\documentclass[iop]{emulateapj}
\usepackage{amssymb}
\usepackage{amsmath}
\usepackage{rotating}
\usepackage{longtable}

\newcommand{\msun}{M_\odot}
\newcommand{\lsun}{L_\odot}

\begin{document}
\title{Variable Stars in Large Magellanic Cloud Globular Clusters III: Reticulum*}
\author{Charles A. Kuehn\altaffilmark{1,2}, Kyra Dame\altaffilmark{1}, Horace A. Smith\altaffilmark{1}, M\'arcio Catelan\altaffilmark{3,4}, Young-Beom Jeon\altaffilmark{5}, James M. Nemec\altaffilmark{6}, Alistair R. Walker\altaffilmark{7}, Andrea Kunder\altaffilmark{7}, Barton J. Pritzl\altaffilmark{8}, Nathan De Lee\altaffilmark{1,9}, Jura Borissova\altaffilmark{10,4}}
\altaffiltext{*}{Based on observations taken with the SMARTS $1.3$-meter telescope operated by the SMARTS Consortium and observations taken at the Southern Astrophysical Research (SOAR) telescope, which is a joint project of the Minist\'{e}rio da Ci\^{e}ncia, Tecnologia, e Inova\c{c}\~{a}o (MCTI) da Rep\'{u}blica Federativa do Brasil, the U.S. National Optical Astronomy Observatory (NOAO), the University of North Carolina at Chapel Hill (UNC), and Michigan State University (MSU).}
\altaffiltext{1}{Department of Physics and Astronomy, Michigan State University, East Lansing, MI 48824, USA; damekyra@msu.edu, smith@pa.msu.edu}
\altaffiltext{2}{Current address: Sydney Institute for Astronomy, University of Sydney, Sydney, Australia; kuehn@physics.usyd.edu.au}
\altaffiltext{3}{Pontificia Universidad Cat$\rm{\acute{o}}$lica de Chile, Facultad de F\'{i}sica, Departamento de Astronom\'ia y Astrof\'isica, Santiago, Chile; mcatelan@astro.puc.cl}
\altaffiltext{4}{The Milky Way Millennium Nucleus, Santiago, Chile}
\altaffiltext{5}{Korea Astronomy and Space Science Institute, Daejeon, Korea; ybjeon@kasi.re.kr}
\altaffiltext{6}{Department of Physics and Astronomy, Camosun College, Victoria, British Columbia, Canada; nemec@camosun.bc.ca}
\altaffiltext{7}{Cerro Tololo Inter-American Observatory, National Optical Astronomy Observatory, La Serena, Chile; awalker@ctio.noao.edu,akunder@ctio.noao.edu}
\altaffiltext{8}{Department of Physics and Astronomy, University of Wisconsin Oskosh, Oshkosh, WI 54901, USA; pritzlb@uwosh.edu}
\altaffiltext{9}{Current address: Department of Physics and Astronomy, Vanderbilt University, Nashville, TN 37235, USA; nathan.delee@vanderbilt.edu}
\altaffiltext{10}{Departamento de F\'isica y Astronom\'ia, Falcultad de Ciencias, Universidad de Valpara\'iso, Valpara\'iso, Chile; jura.borissova@uv.cl}

\begin{abstract}
This is the third in a series of papers studying the variable stars in old globular clusters in the Large Magellanic Cloud.  The primary goal of this series is to look at how the characteristics and behavior of RR Lyrae stars in Oosterhoff-intermediate systems compare to those of their counterparts in Oosterhoff-I/II systems.  In this paper we present the results of our new time-series BVI photometric study of the globular cluster Reticulum.  We found a total of $32$ variables stars ($22$ RRab, $4$ RRc, and $6$ RRd stars) in our field of view.  We present photometric parameters and light curves for these stars.  We also present physical properties, derived from Fourier analysis of light curves, for some of the RR Lyrae stars.  We discuss the Oosterhoff classification of Reticulum and use our results to re-derive the distance modulus and age of the cluster.
\end{abstract}


\section{Introduction}

This is the third in a series of papers focusing on the variable stars in Large Magellanic Cloud (LMC) globular clusters.  The goal of this series of papers is to better understand the nature of the Oosterhoff dichotomy in the Milky Way and how Oosterhoff intermediate (Oo-int) clusters in nearby dwarf galaxies fit into that picture.  Globular clusters in the Milky Way are classified as either Oosterhoff I (Oo-I) or Oosterhoff II (Oo-II) objects based on the properties of their RR Lyrae stars.  Oo-I objects are defined as having an average RRab period of $\langle P_{ab}\rangle < 0.58$ days while Oo-II objects have $\langle P_{ab}\rangle > 0.62$ days; the typical values of $\langle P_{ab}\rangle$ are $0.55$ days and $0.65$ days for Oo-I and Oo-II objects, respectively.  Oo-I clusters also tend to be more metal-rich and have a smaller ratio of first overtone dominant to fundamental mode dominant RR Lyrae. The period range between the two groups, $0.58\leq\langle P_{ab}\rangle \leq 0.62$ days, is referred to as the Oosterhoff gap and is essentially unoccupied by Milky Way globular clusters.  The nearby dwarf galaxies and their globular clusters present a sharp contrast to this behavior as these extra-galactic clusters not only fall in the Oo-I and Oo-II groups, they also fall into the gap between these groups; in fact the extra-galactic objects seem to preferentially be located in the gap \citep{catelan09a}.  These Oosterhoff-intermediate objects, as objects that fall in the gap are called, present a challenge for models that propose that the Milky Way halo was formed through the accretion of objects similar to the present day nearby dwarf galaxies as we would expect to see similar Oosterhoff properties in both samples if that were the case.

The first two papers in this series discussed the variables in the globular clusters NGC 1466 \citep{kuehn11} and NGC 1786 \citep{kuehn12}.  These previous investigations, combined with the results presented here, build an inventory of updated RR Lyrae properties in a representative sample of LMC globular clusters.  A future paper in the series will present a more detailed discussion of our present understanding of the Oosterhoff phenomenon and how the overall results from our study of LMC globular clusters fit into this picture.

Reticulum is an old globular cluster that is located $\approx 11^{\circ}$ from the center of the LMC \citep{demers76}.  It has a metal abundance of ${\rm [Fe/H]_{ZW84}}\approx -1.66$ \citep{mackeygilmore04} (${\rm [Fe/H]_{UVES}}\approx -1.61$ in the new UVES scale \citep{carretta09}) and is not very reddened, E(B-V) = $0.016$ \citep{schlegel98}.  Mackey \& Gilmore found that the age of Reticulum is similar to the ages of the oldest globular clusters in the Milky Way and the LMC, having an age that is approximately $1.4$ Gyr younger than the classic nearby Milky Way halo globular cluster M3.  \citet{johnson02} used Hubble Space Telescope observations to determine that Reticulum formed within $2$ Gyr of the other old LMC clusters.

\begin{figure}
\epsscale{1.3}
\plotone{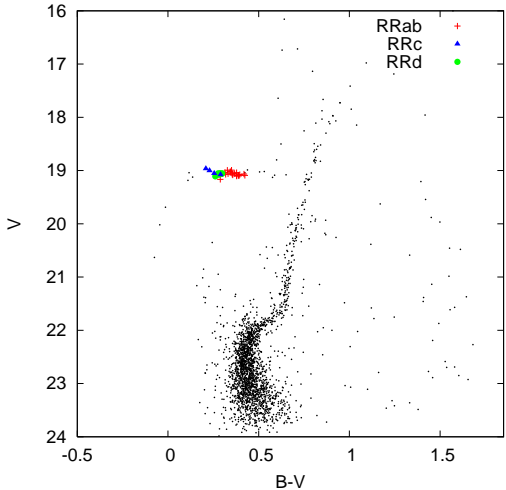}
\caption{$V$,$B-V$ CMD for Reticulum with the position of the RR Lyrae variables also indicated.  Plus symbols indicate RRab stars, filled triangles indicated RRc's, and circles indicate RRd's.}
\label{retcmd}
\end{figure}

\begin{figure}
\epsscale{1.3}
\plotone{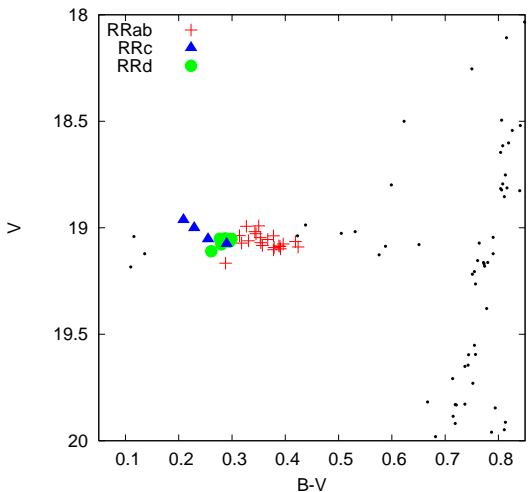}
\caption{$V$,$B-V$ CMD for Reticulum that is zoomed in on the horizontal branch.  The symbols used are the same as in Figure \ref{retcmd}}
\label{retcmdzoom}
\end{figure}

Reticulum is a sparsely populated cluster, but it does have a distinct horizontal branch that stretches across the instability strip (Figures \ref{retcmd} and \ref{retcmdzoom}).  Twenty-two RR Lyrae stars were first found in the cluster by \citet{demers76}.  Walker (1992a, hereafter Walker) later found an additional ten RR Lyrae stars, bringing the total in the cluster to $32$.  The pulsation types include $22$ RRab stars (fundamental-mode pulsators), $9$ RRc's (first-overtone), and $1$ candidate RRd (double-mode pulsators), although recently \citet{ripepi04} found evidence for RRd behavior in four of the previously discovered RR Lyrae stars.  

%No new RR Lyrae stars are expected to be found in the central regions of Reticulum since this area is uncrowded and blending is not an issue.  I obtained $82$ $V$, $72$ $B$, and $78$ $I$ images which comprise a larger data set than the $33$ $BV$ pairs obtained by Walker, allowing me to better sample the light curves of the variable stars.

\section{Observations and Data Analysis}

A total of $38$ $V$, $33$ $B$, and $35$ $I$ images were obtained using the SOI imager ($5.2$x$5.2$ arcminute field of view) on the SOAR $4$-m telescope in February of 2008.  ANDICAM ($6$x$6$ arcminute field of view) on the SMARTS $1.3$-m telescope operated by the SMARTS consortium was used to obtain $45$ $V$, $43$ $B$, and $45$ $I$ images between September 2006 and the end of December 2006.  An additional 145 $V$ and 146 $B$ image were taken with the Tek2K ($13.6$x$13.6$ arcminute field of view) on the SMARTS $0.9$-m telescope between December 2008 and November 2009.  SOAR exposure times were between $30$s and $600$s for $V$ and $I$, and between $45$s and $900$s for $B$.  SMARTS $1.3$-m exposures were $450$s for the $V$ and $B$ filters and $300$s for the $I$ filter.  SMARTS $0.9$-m exposures were 400s in both the $V$ and $B$.

Data reduction and variable identification for the SOAR and SMARTS $1.3$-m data were carried out as described in \citet{kuehn11}, the same method used for both NGC 1466 and NGC 1786.  The SMARTS $0.9$-m data was processed using the method described in \citet{jeon12}.  The uncrowded nature of Reticulum was ideal for Daophot's profile fitting photometry \citep{st87,st92,st94} and while an image differencing method (ISIS; Alard 2000) was run on the images for completeness, no additional variable stars were recovered.  The photometry from Daophot was transformed to the standard system using the Landolt standard fields PG0231, SA95, and SA98 \citep{la92}.  We compared our resulting photometry to five of the local standard stars used by Walker, finding that for these five stars our photometry was $0.011\pm0.010$ magnitudes brighter in $V$ and $0.001\pm0.018$ in $B$.

\section{Variable Stars}

\begin{figure*}
\includegraphics[width=0.45\textwidth]{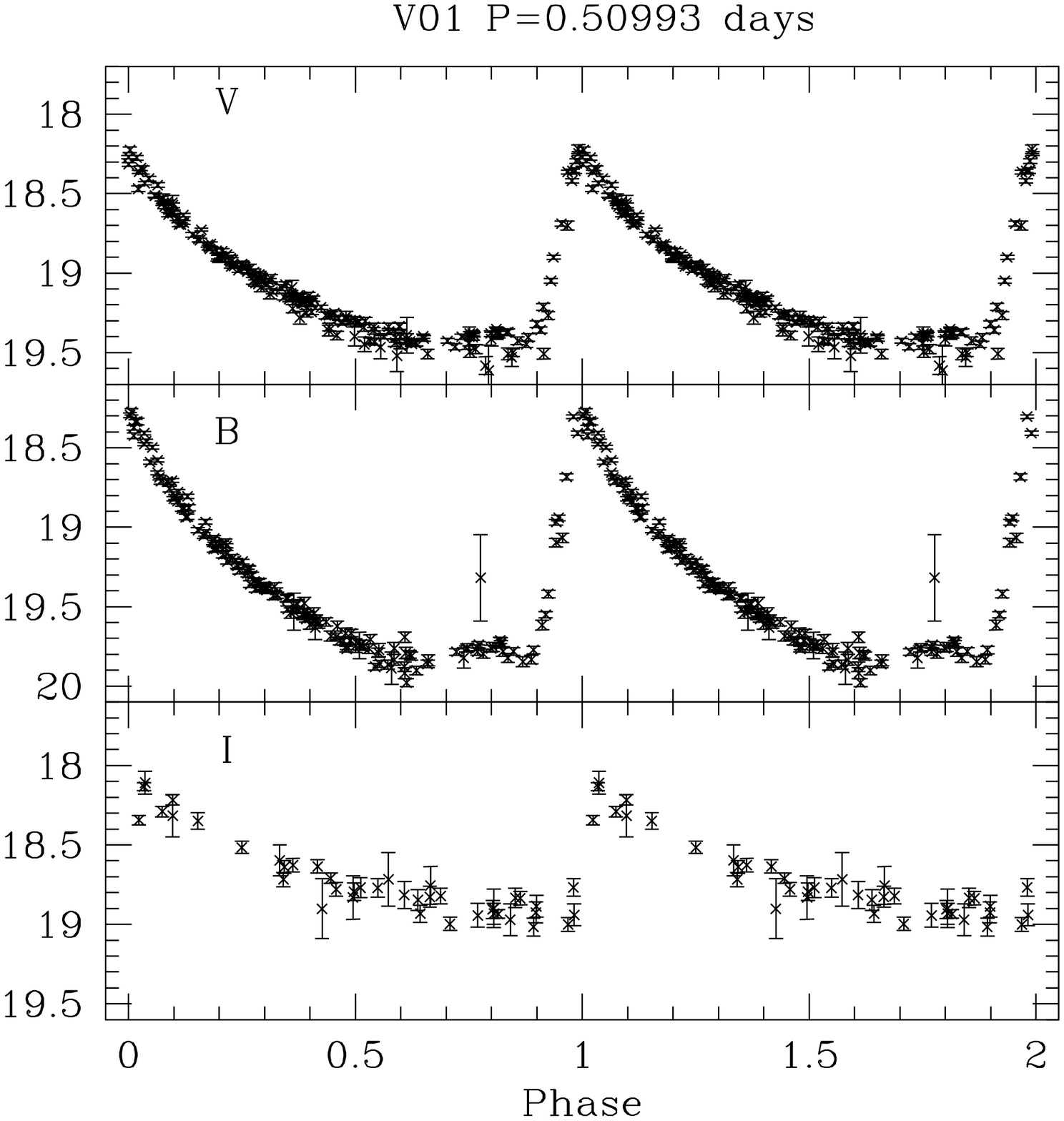}
\includegraphics[width=0.45\textwidth]{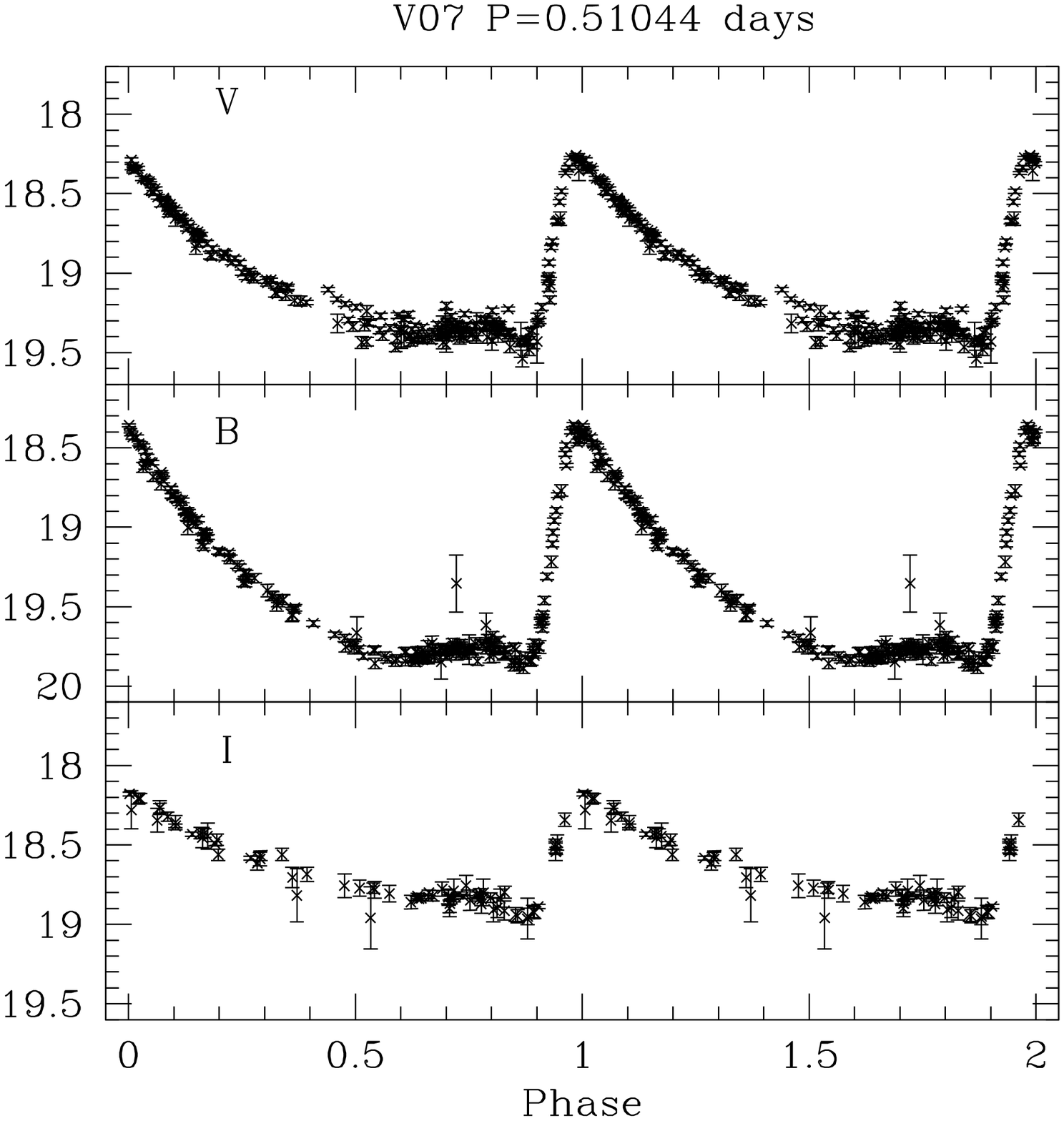}
\caption{Sample light curves for RRab stars in Reticulum, the full set of light curves can be found in the electronic version of this paper.}
\label{retab}
\end{figure*}

\begin{figure*}
\includegraphics[width=0.45\textwidth]{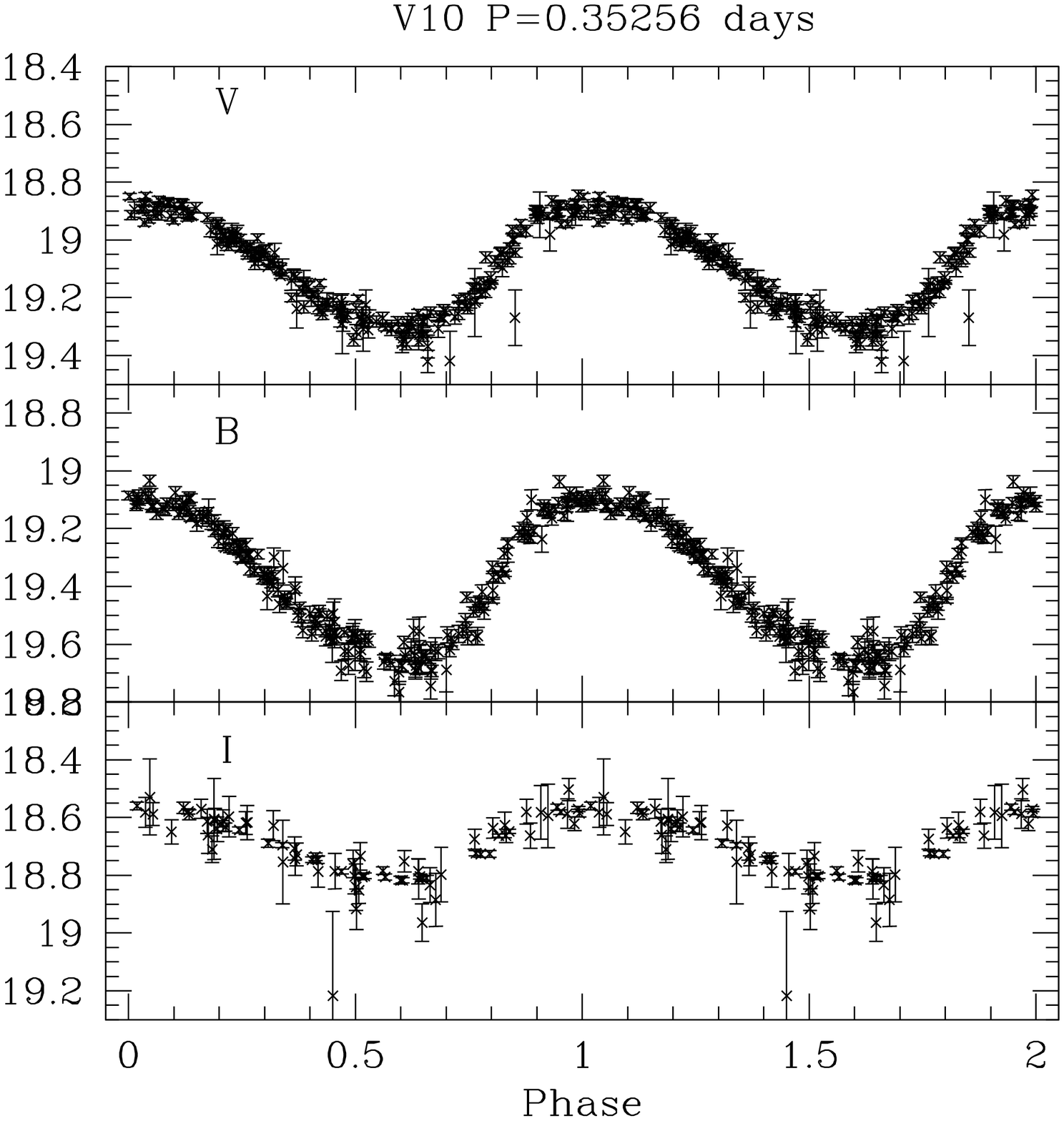}
\includegraphics[width=0.45\textwidth]{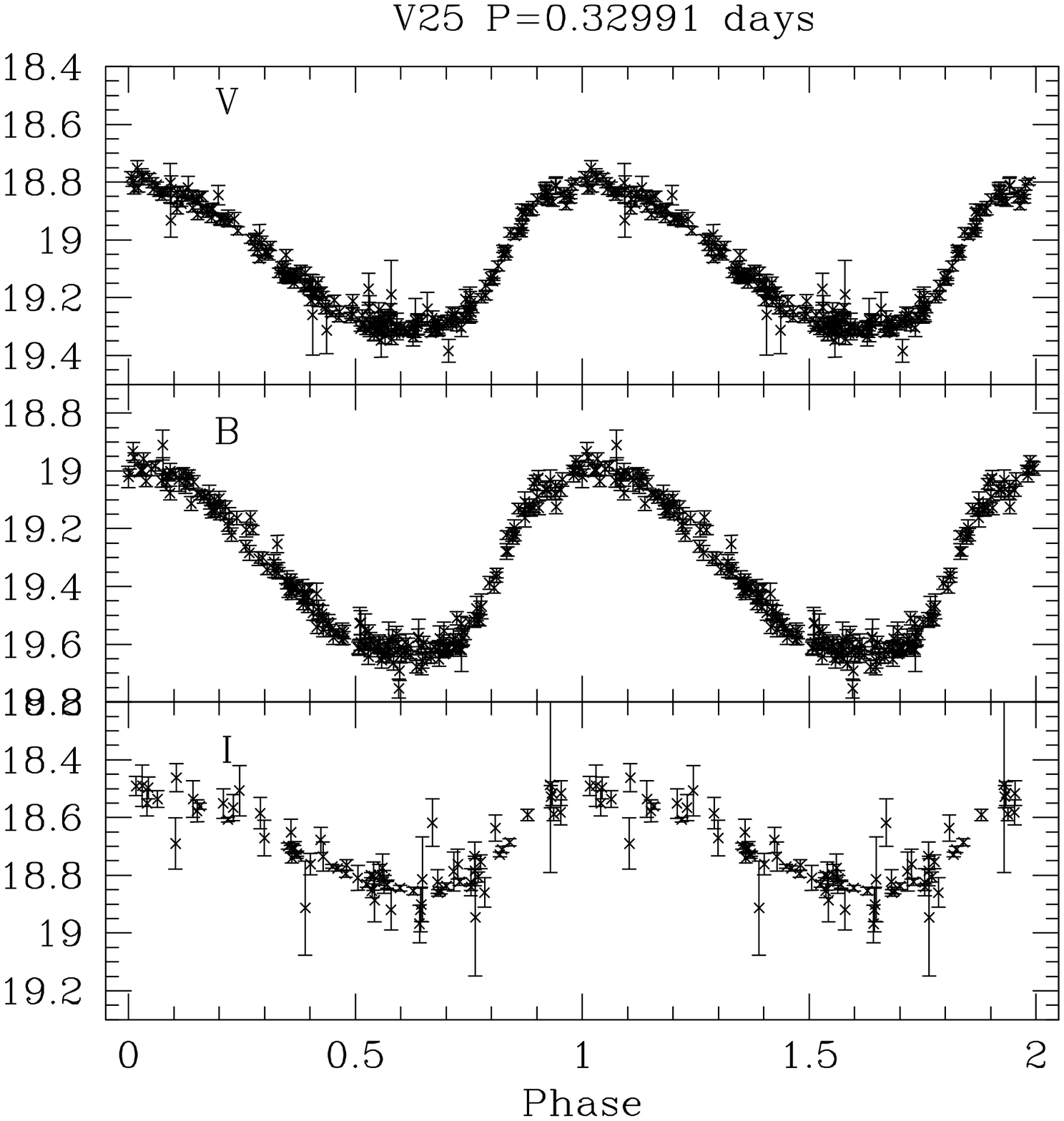}
\caption{Sample light curves for RRc stars in Reticulum, the full set of light curves can be found in the electronic version of this paper.}
\label{retc}
\end{figure*}

\begin{figure*}
\includegraphics[width=0.45\textwidth]{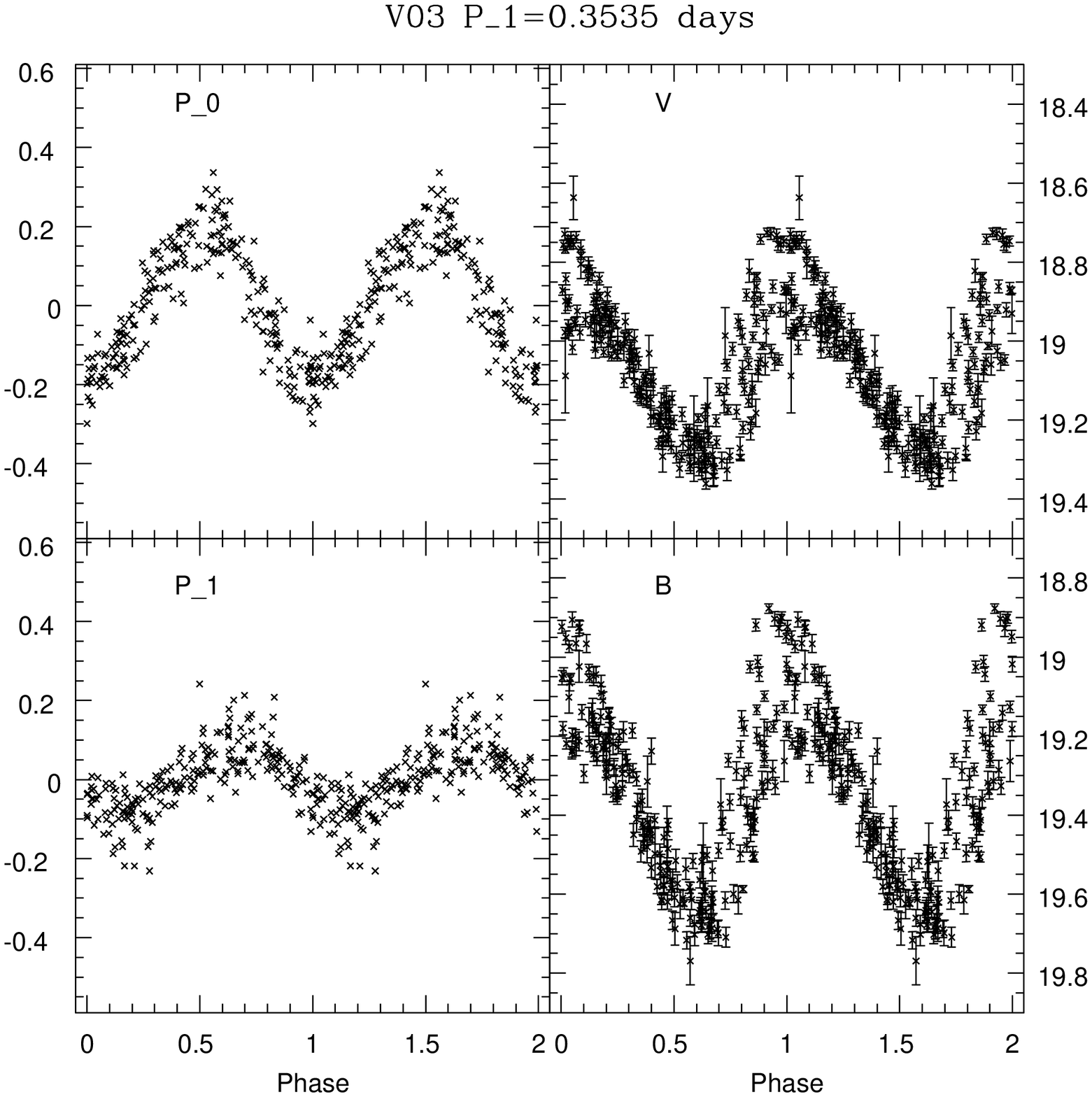}
\includegraphics[width=0.45\textwidth]{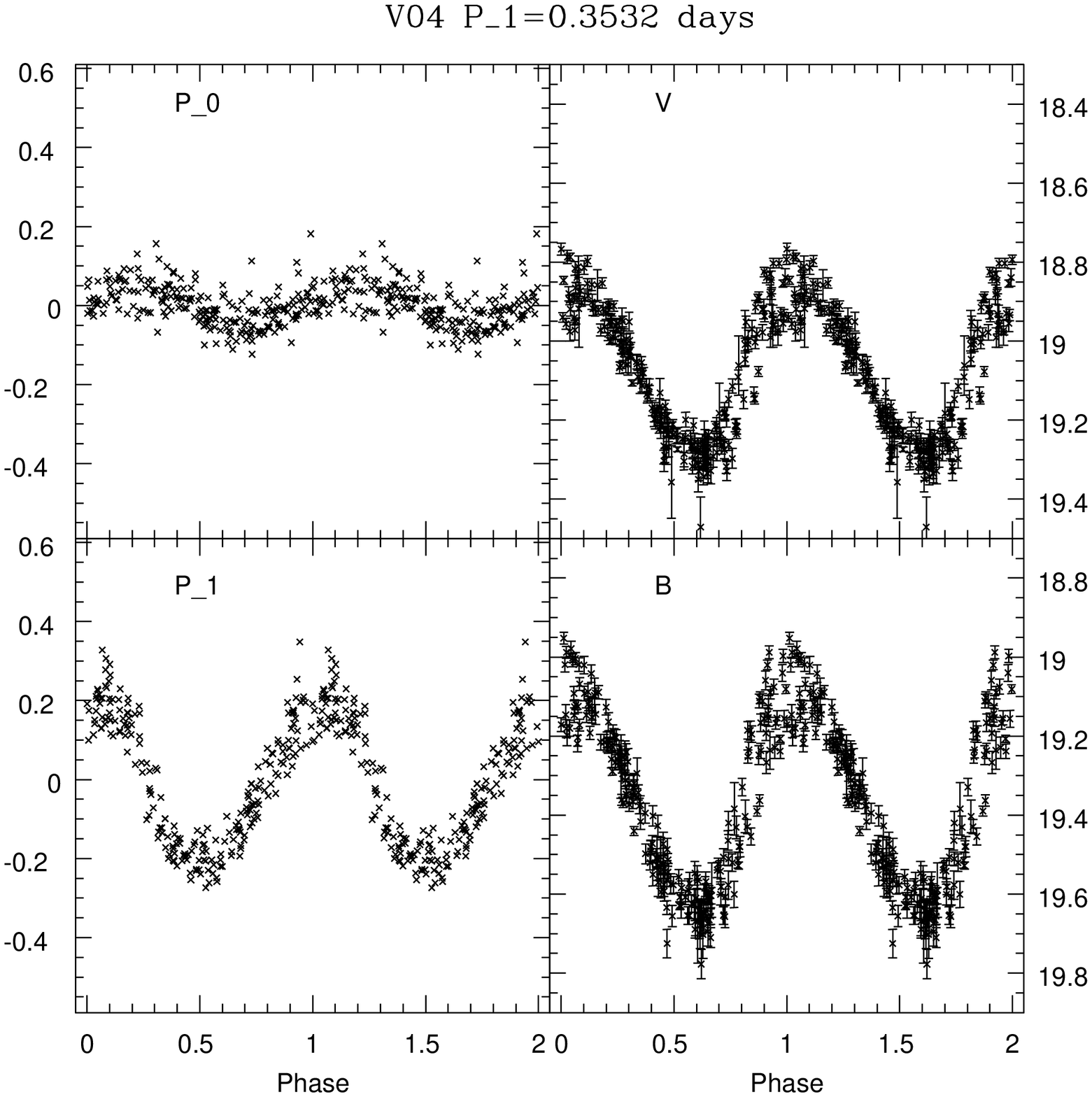}
\caption{Sample light curves for RRd stars in Reticulum.  The left panels show the residuals in the $V$-band light curves left from subtracting the fundamental or first overtone periods.  The right panels show the $V$ and $B$-band light curves plotted with the first overtone periods.  The full set of light curves can be found in the electronic version of this paper.}
\label{retd}
\end{figure*}

\begin{figure*}
\epsscale{0.7}
\plotone{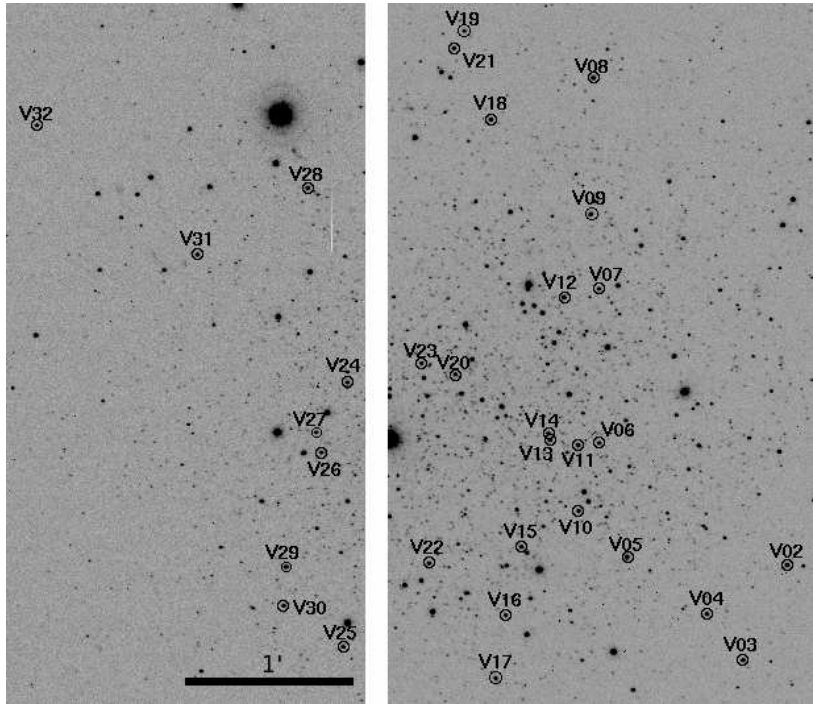}
\caption{Finding chart for the variable stars in Reticulum.  North is up and East is to the left.  The white gap is due to the finding chart being made from a SOAR image and represents the $7.8$ arcsec mounting gap between the two CCDs of the SOI camera.}
\label{retfind}
\end{figure*}

All $32$ RR Lyrae stars found by Walker were recovered: $22$ RRab stars, $4$ RRc's, and $6$ RRd's.  The $6$ RRd stars were originally classified as RRc stars by Walker but the larger number of observations in our data set allowed for the identification of secondary pulsation modes.  The RRab and RRc stars and their observed characteristics (periods, $V$, $B$, and $I$ amplitudes, intensity-weighted $V$, $B$, and $I$ mean magnitudes, and magnitude-weighted mean $B-V$ color) are listed in Table \ref{reticulumvartable}; the stars that potentially show the Blazhko effect are identified with 'BL' after their name.  The RRd stars, their fundamental and first overtone periods and amplitudes, their period ratios, and their mean magnitudes and color are listed in Table \ref{reticulumrrd}.  Periods for RRab and RRc stars are typically good to $\pm0.00001$ or $\pm0.00002$ days while periods for RRd stars are less well known, with uncertainties about an order of magnitude larger.  Walker identified the variables in his paper using their star number in the catalog compiled in \citet{demers76}.  We introduce a new naming system that features only the variable stars and is ordered based on increasing RA.  The names used by Walker are listed in the last column in Tables \ref{reticulumvartable} and \ref{reticulumrrd}.  Table \ref{phot} gives the photometry for the RR Lyrae stars and Figures \ref{retab}, \ref{retc}, and \ref{retd} show the light curves for the RRab, RRc, and RRd stars, respectively.  The positions of the variable stars within the cluster are shown in Figure \ref{retfind}.

\begin{deluxetable*}{lcccccccccccc}
\tablewidth{0pc}
\tabletypesize{\scriptsize}
\tablecaption{Photometric Parameters for Variables in Reticulum}
\tablehead{\colhead{ID}& \colhead{RA (J2000)} & \colhead{DEC (J2000)} & \colhead{Type}& \colhead{$P$ (days)}& \colhead{$A_{V}$}& \colhead{$A_{B}$}& \colhead{$A_{I}$}& \colhead{$\langle V\rangle$}& \colhead{$\langle B\rangle$}& \colhead{$\langle I\rangle$}& \colhead{$\langle B-V\rangle$} & \colhead{Other IDs}}
\startdata
V01&   04:35:51.4&  -58:51:03.4&  RRab&  0.50993&  1.21& 1.62&  0.85&	19.037&  19.299& 18.651&	0.314&  DK92\\
V02&   04:35:56.5&  -58:52:32.0&  RRab&  0.61869&  0.66& 0.91&  0.44&	19.085&  19.455& 18.561&	0.388&  DK7\\
V05&   04:36:04.1&  -58:52:29.4&  RRab&  0.57185&  0.94& 1.28&  0.55&	19.038&  19.386& 18.606&	0.378&  DK80\\
V06-BL&   04:36:05.4&  -58:51:48.0&  RRab&  0.59526&  0.84& 1.10&  0.58&	19.099&  19.476& 18.583&	0.391&  DK97\\
V07&   04:36:05.5&  -58:50:51.9&  RRab&  0.51044&  1.18& 1.56&  0.75&	18.991&  19.298& 18.638&	0.350&  DK117\\
V08&   04:36:05.8&  -58:49:35.1&  RRab&  0.64496&  0.45& 0.59&  0.26&	19.065&  19.475& 18.550&	0.419&  DK135\\
V09&   04:36:05.9&  -58:50:24.6&  RRab&  0.54496&  0.92& 1.08&  0.45&	18.993&  19.295& 18.589&	0.327&  DK137\\
V10&   04:36:06.4&  -58:52:12.8&  RRc &  0.35256&  0.43& 0.56&  0.25&	19.075&  19.358& 18.685&	0.290&  DK77\\
V12&   04:36:07.1&  -58:50:55.0&  RRc &  0.29627&  0.25& 0.31&  0.15&	18.963&  19.169& 18.724&	0.209&  DK181\\
V13&   04:36:07.7&  -58:51:47.0&  RRab&  0.60958&  0.75& 0.94&  0.48&	19.103&  19.462& 18.580&	0.378&  DK99\\
V14-BL&   04:36:07.8&  -58:51:44.6&  RRab&  0.58661&  0.74& 0.94&  0.51&	19.076&  19.451& 18.594&	0.396&  DK100\\
V16&   04:36:09.8&  -58:52:50.8&  RRab&  0.52290&  1.21& 1.61&  0.79&	19.046&  19.351& 18.629&	0.353&  DK49\\
V17&   04:36:10.2&  -58:53:13.7&  RRab&  0.51241&  1.15& 1.45&  0.70&	19.028&  19.325& 18.650&	0.345&  DK38\\
V18&   04:36:10.6&  -58:49:50.4&  RRab&  0.56005&  1.00& 1.34&  0.61&	19.083&  19.405& 18.615&	0.357&  DK142\\
V19&   04:36:11.9&  -58:49:18.2&  RRab&  0.48485&  1.32& 1.73&  0.78&	19.057&  19.303& 18.682&	0.298&  DK146\\
V20-BL&   04:36:12.2&  -58:51:23.3&  RRab&  0.56075&  1.03& 1.36&  0.59&	19.088&  19.455& 18.645&	0.390&  DK112\\
V21&   04:36:12.3&  -58:49:24.5&  RRab&  0.60700&  0.72& 0.95&  0.52&	19.093&  19.452& 18.564&	0.379&  DK145\\
V22&   04:36:13.4&  -58:52:32.1&  RRab&  0.51359&  0.97& 1.21&  0.58&	19.072&  19.355& 18.620&	0.318&  DK57\\
V23-BL&   04:36:13.8&  -58:51:19.3&  RRab&  0.46863&  0.96& 1.30&  0.59&	19.166&  19.415& 18.740&	0.288&  DK108\\
V25&   04:36:17.4&  -58:53:02.7&  RRc &  0.32991&  0.52& 0.67&  0.33&	19.053&  19.297& 18.681&	0.255&  DK36\\
V26&   04:36:18.5&  -58:51:52.0&  RRab&  0.65696&  0.34& 0.46&  0.22& 	19.090&  19.509& 18.553&	0.424&  DK67\\
V27&   04:36:18.7&  -58:51:44.6&  RRab&  0.51382&  1.25& 1.64&  0.79&	19.055&  19.374& 18.637&	0.367&  DK64\\
V28&   04:36:19.2&  -58:50:15.7&  RRc &  0.31994&  0.51& 0.66&  0.32&	19.001&  19.219& 18.679&	0.229&  DK151\\
V29&   04:36:20.1&  -58:52:33.6&  RRab&  0.50815&  1.21& 1.58&  0.80&	19.061&  19.338& 18.640&	0.331&  DK37\\
V30&   04:36:20.2&  -58:52:47.7&  RRab&  0.53501&  1.16& 1.52&  0.72&	19.018&  19.318& 18.577&	0.343&  DK35\\
V31&   04:36:24.4&  -58:50:40.1&  RRab&  0.50516&  1.11& 1.52&  0.67&	19.070&  19.379& 18.681&	0.354&  DK25\\
\enddata
\label{reticulumvartable}
\end{deluxetable*}

\begin{deluxetable*}{lcccccccccccccccc}
%\rotate{90}
\tablewidth{0pc}
\tabletypesize{\scriptsize}
\tablecaption{Photometric Parameters for the RRd Variables in Reticulum}
\tablehead{\colhead{ID} & \colhead{RA (J2000)} & \colhead{DEC (J2000)} & \colhead{$P_{0}$ (d)} & \colhead{$P_{1}$ (d)} & \colhead{$P_{1}/P_{0}$} & \colhead{$A_{V,0}$} & \colhead{$A_{V,1}$} & \colhead{$A_{B,0}$} & \colhead{$A_{B,1}$} & \colhead{$A_{I,0}$} & \colhead{$A_{I,1}$} & \colhead{$\langle V\rangle $} & \colhead{$\langle B\rangle $} & \colhead{$\langle I\rangle $} & \colhead{$\langle B-V\rangle$} & \colhead{Other IDs}}
\startdata
V03&   04:35:58.6&   -58:53:06.7&      0.4751&	0.3535&	 0.7442& 0.21&  0.41&  	0.25&	0.53&   0.13&	0.25&	19.050&   19.329&   18.662&	0.287&   DK41\\
V04&   04:36:00.3&   -58:52:50.0&      0.4747&	0.3532&	 0.7440& 0.12&	0.43&	0.13&	0.56&   0.13&	0.30&	19.065&   19.351&   18.674&	0.295&   DK4\\
V11&   04:36:06.4&   -58:51:48.7&      0.4777&  	0.3554&  0.7439& 0.32&	0.41&  	0.44& 	0.55&	0.23&	0.26&	19.052&	  19.341&   18.693&	0.299&   DK98\\
V15&   04:36:09.1&   -58:52:25.9&      0.4761&	0.3543&	 0.7441& 0.28&	0.41&	0.38&	0.53&   0.19&	0.21&	19.110&   19.364&   18.721&	0.261&   DK72\\
V24&   04:36:17.3&   -58:51:26.3&      0.4670&	0.3475&	 0.7441& 0.27&	0.44&	0.26&	0.52&   0.17&	0.24&	19.078&   19.350&   18.673&	0.280&   DK110\\
V32&   04:36:31.9&   -58:49:53.2&      0.4734&	0.3523&  0.7441& 0.12&	0.42&	0.10&	0.54&   0.05&	0.24&	19.052&   19.322&   18.649&	0.277&   DK157\\
\enddata
\label{reticulumrrd}
\end{deluxetable*}

\begin{deluxetable*}{lcccccc}
\tablewidth{0pc}
\tabletypesize{\scriptsize}
\tablecaption{Photometry of the Variable Stars}
\tablehead{\colhead{ID} & \colhead{Filter} & \colhead{JD}  & \colhead{Phase} & \colhead{Mag} & \colhead{Mag Error} & \colhead{Telescope} }
\startdata
V01 & $B$ & 2453990.7614 & $0.95651$ & $19.067$ & $0.029$ & $1.3$-m\\
V01 & $B$ & 2454004.7307 & $0.35079$ & $19.515$ & $0.020$ & $1.3$-m\\
V01 & $B$ & 2454018.7159 & $0.77625$ & $19.318$ & $0.273$ & $1.3$-m\\
V01 & $B$ & 2454023.7373 & $0.62339$ & $19.810$ & $0.026$ & $1.3$-m\\
V01 & $B$ & 2454041.6440 & $0.73904$ & $19.823$ & $0.065$ & $1.3$-m\\
\enddata
\tablecomments{Maximum light occurs at a phase of $0$.  This table is published in its entirety in the electronic edition.}
\label{phot}
\end{deluxetable*}

The RRab stars have intensity-weighted mean magnitudes of $\langle V\rangle = 19.06\pm0.01$, $\langle B\rangle = 19.39\pm0.01$, and $\langle I\rangle=18.61\pm0.01$ while the RRc stars have mean magnitudes of $\langle V\rangle = 19.05\pm0.02$, $\langle B\rangle = 19.26\pm0.04$, and $\langle I\rangle=18.69\pm0.01$.  The results for RRab stars are $0.02$ mag brighter than the mean magnitudes found by Walker while our values for the RRc stars are consistent within the errors of those found by Walker.

%The results for RRab stars are slightly brighter than the mean magnitudes of $\langle V\rangle = 19.08\pm0.01$ and $\langle B\rangle = 19.41\pm0.02$ found by Walker, while ur values for the RRc stars are consistent within the errors to the mean magnitudes of $\langle V\rangle = 19.05\pm0.02$ and $\langle B\rangle = 19.28\pm0.04$ found by Walker.  

%These differences in brightness for the RR Lyrae stars are on the same order as, and are likely due to, the previously mentioned slight photometric disagreement between our results and those of Walker.

In general our periods agreed with those of Walker to within $0.0002$ days.  V$08$ and V$19$ were the only stars for which a difference in period greater than $0.01$ days was found.  Walker found a period of $0.6566$ days for V$08$ while we found a period of $0.64495$ days, a decrease of $0.0117$ days.  For V$19$, Walker found a period of $0.469$ days while we found a period of $0.48485$ days, an increase of $0.016$ days.  We believe the periods adopted represent these stars more accurately, as our phase coverage and timespan of observation is significantly improved over those of Walker.

%These differences in period could be due to either an error in period determination in the previous studies or due to actual period changes.  Period changes in RR Lyrae stars can be used to study the evolution of these stars as they move through the instability strip \citep{catelan09a}.  A decrease in period indicates a blueward movement on the HR diagram while an increase in period indicates a redward move.  In addition to period changes due to evolution, RR Lyrae stars are also known to undergo irregular period changes, of which the exact cause is not known.  The size of these period changes is larger than expected for evolutionary period change and larger than seen in other RR Lyrae, and are thus most likely due to an error in period determination by Walker.

\begin{figure*}
\includegraphics[width=0.45\textwidth]{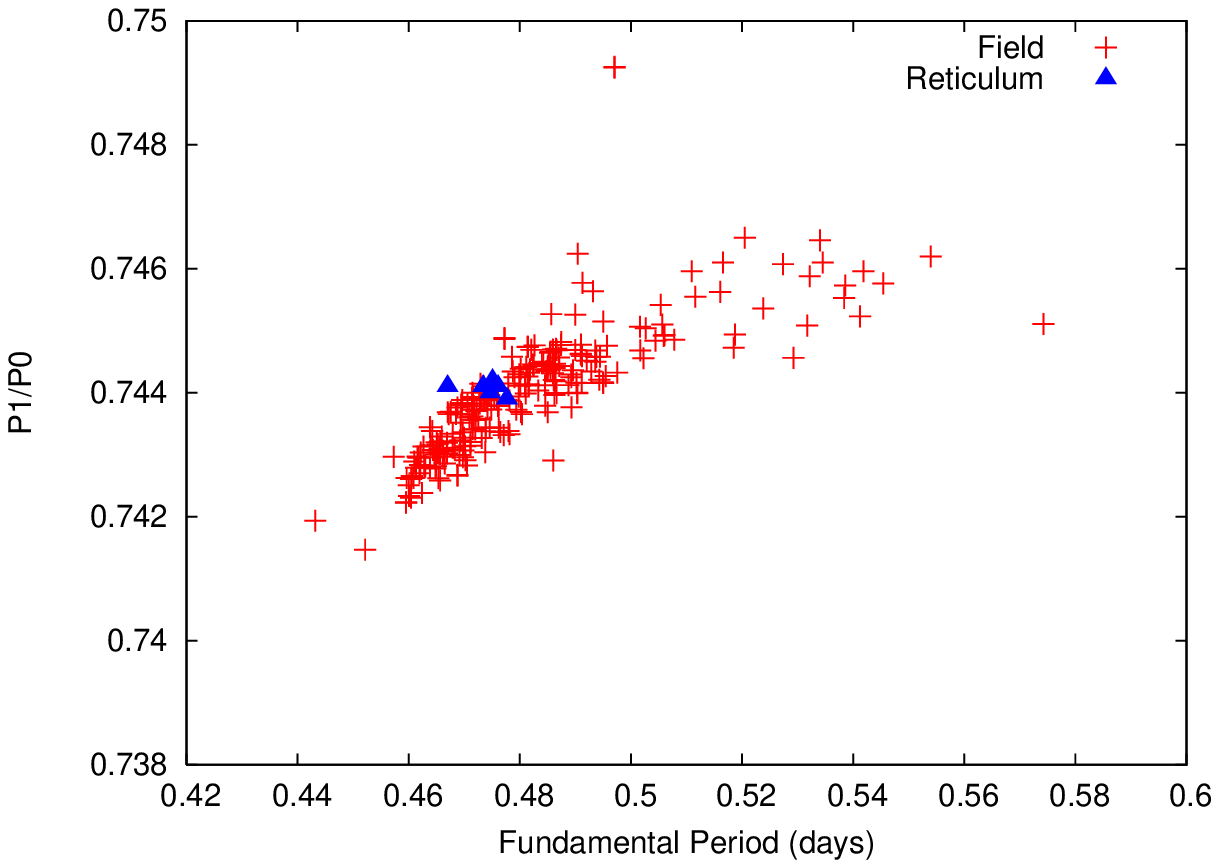}
\includegraphics[width=0.45\textwidth]{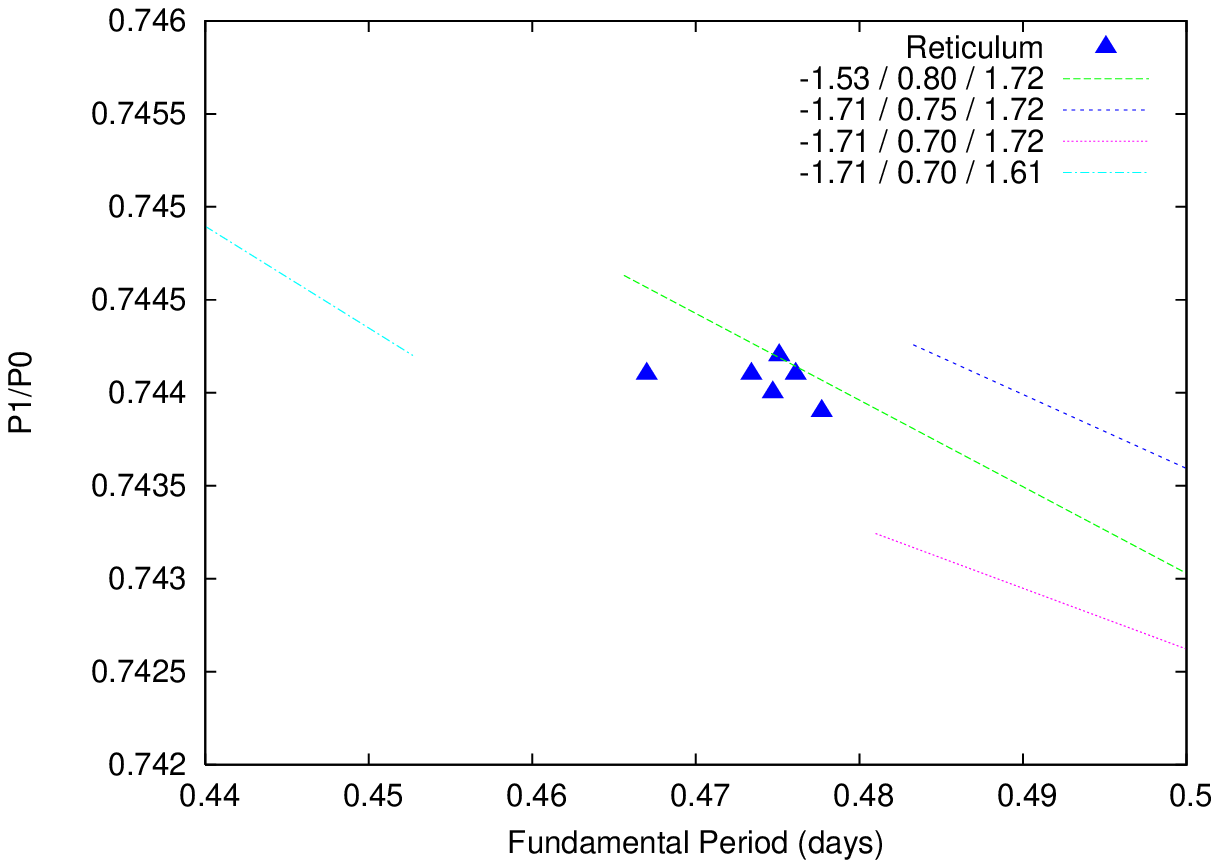}
\caption{Petersen diagram showing the ratio of the first overtone period to the fundamental mode period vs. fundamental mode period for the RRd stars in Reticulum (blue triangles).  (Left panel) The RRd stars in the LMC field (red plus symbols) from \citet{soszy03} are also plotted.  (Right panel)  The colored lines are from the models by \citet{bragaglia01}; their labels indicate the assumed metallicity ($\rm {[Fe/H]_{ZW84}}$), mass ratio ($M/\msun$), and luminosity ($\log(L/\lsun)$).}
\label{retpetersen}
\end{figure*}

The first overtone periods for the RRd stars show good agreement with the periods that Walker had reported.  Four of the six RRd stars (V$03$, V$11$, V$15$, V$24$) were also found by \citet{ripepi04}.  Figure \ref{retpetersen} shows the Petersen diagram for the RRd stars in Reticulum along with those in the field of the LMC.  The Reticulum RRd stars have similar period ratios to not only the LMC field RRd stars, but also to the RRd stars found in Milky Way Oo-I clusters \citep{clementini04}.  The right hand panel shows the results for models from \citet{bragaglia01} for three different combinations of metallicity, mass, and luminosity of the RRd stars.  The RRd stars in Reticulum are fit very well by the line that corresponds to a metallicity of ${\rm [Fe/H]_{ZW84}}=-1.53$ (${\rm [Fe/H]_{UVES}}=-1.45$), a mass of $M/\msun=0.80$, and a luminosity of $log(L/\lsun)=1.72$.  However, this gives a mass that is much higher than the masses of the RRc stars calculated through the Fourier decomposition method (see \S4) or from fitting horizontal branch evolutionary tracks to the color-magnitude diagram (\S5).  The model tracks for a metallicity of ${\rm [Fe/H]_{ZW84}}=-1.71$ (${\rm [Fe/H]_{UVES}}=-1.68$) suggests that the Reticulum RRd stars could also be fit by a model of that metallicity with a mass in the range of $0.70 < M/\msun < 0.75$ which would be closer to the masses obtained for the RRc stars.

\section{Physical Properties of the RR Lyrae Stars}

It has been shown that the Fourier parameters of RR Lyrae light curves can be used to estimate their physical properties (e.g., Jurcsik \& Kovacs 1996, Jurcsik 1998, Simon \& Clement 1993).  The RRab light curves were fit with a Fourier series of the form
\begin{equation}
m(t)=A_{0}+\sum_{j=1}^{n}A_{j}\sin (j\omega t+\phi_{j}),
\end{equation}
while the RRc light curves were fit with a cosine series.  The resulting Fourier coefficients were then used to calculate physical properties of the stars using the relations from \citet{jurcsikkovacs96}, \citet{jurcsik98}, Kov\'acs \& Walker (1999, 2001), \citet{simonclement93}, and \citet{morgan07}.  We refer the reader to the first paper in this series, \citet{kuehn11}, for further details. 

\begin{deluxetable*}{lccccccccc}
\tablewidth{0pc}
\tabletypesize{\scriptsize}
\tablecaption{Fourier Coefficients for RRab Variables}
\tablehead{\colhead{ID} & \colhead{$A_{1}$} & \colhead{$A_{21}$} & \colhead{$A_{31}$} & \colhead{$A_{41}$} & \colhead{$\phi_{21}$} & \colhead{$\phi_{31}$} & \colhead{$\phi_{41}$}& \colhead{$D_{max}$}& \colhead{Order}}
\startdata
V02&	0.243&	0.455&	0.306&	0.166&	2.469&	5.152$\pm$0.051&	1.817&	2.0&  7\\
V05&	0.320&	0.460&	0.338&	0.235&	2.315&	4.972$\pm$0.043&	1.356&	1.6&  7\\
V13&	0.267&	0.462&	0.330&	0.204&	2.375&	5.130$\pm$0.053&	1.632&	3.6&  7\\
V18&	0.339&	0.453&	0.359&	0.227&	2.286&	4.884$\pm$0.045&	1.349&	4.3&  8\\
V21&	0.259&	0.453&	0.325&	0.209&	2.438&	5.166$\pm$0.075&	1.772&	3.2&  9\\
\hline
\\
V07&	0.404&	0.457&	0.330&	0.248&	2.234&	4.694$\pm$0.056&	0.933&	42.1& 8\\
V14&	0.273&	0.449&	0.336&	0.208&	2.373&	4.940$\pm$0.091&	1.334&	46.4& 8\\
V16&	0.416&	0.426&	0.324&	0.237&	2.273&	4.777$\pm$0.041&	1.090&	44.0& 9\\
V17&	0.390&	0.425&	0.330&	0.207&	2.191&	4.625$\pm$0.056&	1.027&	44.5& 7\\
V22&	0.349&	0.443&	0.308&	0.155&	2.332&	4.754$\pm$0.062&	1.075&	40.6& 8\\
V27&	0.433&	0.446&	0.352&	0.224&	2.277&	4.700$\pm$0.036&	1.061&	45.4& 9\\
V29&	0.416&	0.468&	0.325&	0.221&	2.217&	4.676$\pm$0.045&	0.993&	43.8& 8\\
V30&	0.400&	0.447&	0.323&	0.196&	2.308&	4.922$\pm$0.063&	1.151&	40.9& 8\\
V31&	0.400&	0.428&	0.335&	0.208&	2.237&	4.755$\pm$0.044&	1.007&	43.0& 8\\
\enddata
\label{retabcoeff}
\end{deluxetable*}

\begin{deluxetable*}{lcccccccc}
\tablewidth{0pc}
\tabletypesize{\scriptsize}
\tablecaption{Fourier Coefficients for RRc Variables}
\tablehead{\colhead{ID}&  \colhead{$A_{1}$}& \colhead{$A_{21}$}& \colhead{$A_{31}$}& \colhead{$A_{41}$}& \colhead{$\phi_{21}$}& \colhead{$\phi_{31}$}& \colhead{$\phi_{41}$}& \colhead{Order}}
\startdata
V10&	0.219&	0.134&	0.076&	0.031&	4.878&	3.467$\pm$0.207&	2.244&	6\\
V12&    0.127&  0.135&  0.030&  0.021&  4.161&  1.495$\pm$1.037&        1.856&  6\\
V25&	0.247&	0.135&	0.086&	0.045&	4.822&	3.258$\pm$0.143&	1.715&	7\\
V28&	0.245&	0.170&	0.097&	0.034&	4.712&	3.023$\pm$0.224&	0.898&	6\\
\enddata
\label{retccoeff}
\end{deluxetable*}

Although a Fourier decomposition was attempted on all the RRab and RRc stars, only $14$ RRab had light curves that allowed the reliable determination of Fourier parameters; all $4$ RRc stars had reliable parameters determined but V$12$ stands out from the other RRc stars, see discussion below.  Tables \ref{retabcoeff} and \ref{retccoeff} give the Fourier coefficients for the RRab and RRc stars, respectively.  The physical properties determined from these coefficients are given in Tables \ref{retabphysical} and \ref{retcphysical}.  Table \ref{retabcoeff} also lists the Jurcsik \& Kov\'acs $D_{max}$ values  \citep{jurcsikkovacs96} for the RRab stars.  $D_{max}$ can be used to separate RRab stars with ``regular'' light curves from those with more ``anomalous'' light curves; lower values represent more regular light curves.  \citet{jurcsikkovacs96} suggest that stars with $D_{max} > 3$ should not be trusted to provide reliable physical properties.  We take a slightly more liberal approach and use the RRab stars with $D_{max} < 5$ to determine the average properties for the cluster; following the condition from Jurcsik \& Kov\'acs does not change the average values by a significant amount.

The mean metallicity of the RRab stars is ${\rm [Fe/H]_{J95}}=-1.43\pm0.02$ which is ${\rm [Fe/H]_{ZW84}}=-1.61\pm0.02$ on the Zinn \& West scale and ${\rm [Fe/H]_{UVES}}=-1.55\pm0.04$  This value is in similar to the metallicity of ${\rm [Fe/H]_{UVES}} \simeq -1.61$ found by \citet{mackeygilmore04}.  On the other hand, the relation from \citet{morgan07} gives a metallicity for the RRc stars of ${\rm [Fe/H]_{ZW84}}=-1.75\pm0.03$, ${\rm [Fe/H]_{UVES}}=-1.73\pm0.06$, which is more metal-poor than the values obtained from the RRab stars and the literature, but still within the 0.2 dex error estimation in their empirical relation.  While this difference in metallicity could be caused by errors in the Fourier analysis, the fact that the other physical properties obtained for the RRc stars are consistent with expectations lends support to the validity of the obtained Fourier coefficients.

The Fourier parameters and physical properties for the RRc star V$12$ are listed in Tables \ref{retccoeff} and \ref{retcphysical} but were not included when calculating the average physical properties for the RRc stars in Reticulum.  The physical properties calculated for V$12$ show a marked difference from the properties of the other RRc stars in the cluster; we obtained a metallicity of ${\rm [Fe/H]_{ZW84}}=-1.98$ for V$12$, more metal poor than the other RRcs, and a mass of $M/\msun=0.89$ which is very large for an RRc.  The amplitudes of V$12$ in all three filters are significantly smaller than for the other RRc stars in the cluster, suggesting possible blending.  V$12$ is slightly brighter in $V$ and $B$ and is bluer than the other RRc stars in the cluster, supporting the possibility that it is blended with a faint blue companion.  Blending with a nearby star would alter the shape of the light curve and thus explain the unusual values obtained for the physical properties of V$12$.

\begin{deluxetable*}{lccccccccc}
\tablewidth{0pc}
\tabletypesize{\scriptsize}
\tablecaption{Derived Physical Properties for RRab Variables}
\tablehead{\colhead{ID}& \colhead{${\rm [Fe/H]_{J95}}$}& \colhead{$\langle M_{V}\rangle$}& \colhead{$\langle V-K\rangle$}& \colhead{$\log T_{\rm eff}^{\langle V-K\rangle}$}& \colhead{$\langle B-V\rangle$}& \colhead{$\log T_{\rm eff}^{\langle B-V\rangle}$}& \colhead{$\langle V-I\rangle$}& \colhead{$\log T_{\rm eff}^{\langle V-I\rangle}$}& \colhead{$\log g$}}
\startdata
V02&  -1.446& 0.772& 1.204& 3.800& 0.366& 3.802& 0.528& 3.801& 2.740\\
V05&  -1.435& 0.782& 1.137& 3.807& 0.346& 3.809& 0.502& 3.808& 2.780\\
V13&  -1.427& 0.771& 1.179& 3.803& 0.362& 3.804& 0.522& 3.803& 2.747\\
V18&  -1.490& 0.780& 1.137& 3.807& 0.342& 3.810& 0.497& 3.810& 2.791\\
V21&  -1.363& 0.782& 1.178& 3.802& 0.363& 3.804& 0.523& 3.803& 2.749\\
\hline
\\
V07&  -1.478& 0.799& 1.076& 3.814& 0.317& 3.819& 0.466& 3.818& 2.840\\
V14&  -1.557& 0.781& 1.175& 3.803& 0.358& 3.804& 0.517& 3.805& 2.767\\
V16&  -1.433& 0.785& 1.078& 3.814& 0.316& 3.819& 0.464& 3.818& 2.827\\
V17&  -1.581& 0.796& 1.104& 3.811& 0.321& 3.817& 0.470& 3.817& 2.838\\
V22&  -1.414& 0.827& 1.089& 3.812& 0.328& 3.816& 0.478& 3.815& 2.837\\
V27&  -1.489& 0.781& 1.079& 3.814& 0.314& 3.820& 0.462& 3.819& 2.836\\
V29&  -1.489& 0.795& 1.077& 3.814& 0.314& 3.820& 0.461& 3.819& 2.842\\
V30&  -1.303& 0.790& 1.068& 3.815& 0.321& 3.818& 0.471& 3.816& 2.815\\
V31&  -1.368& 0.815& 1.060& 3.816& 0.318& 3.820& 0.467& 3.817& 2.845\\
\hline
\rm{Mean}& -1.432$\pm$0.020& 0.778$\pm$0.003& 1.167$\pm$0.013& 3.804$\pm$0.001& 0.356$\pm$0.005& 3.806$\pm$0.002& 0.514$\pm$0.006& 3.805$\pm$0.002& 2.762$\pm$0.010\\
\enddata
\tablecomments{The properties in this table were calculated from the Fourier coefficients for the light curves using the equations described in \citet{kuehn11}.  The mean values were computed using only the first five stars, which have the lowest $D_{max}$ values (Table \ref{retabcoeff}).} 
\label{retabphysical}
\end{deluxetable*}

\begin{deluxetable*}{lcccccc}
\tablewidth{0pc}
\tabletypesize{\scriptsize}
\tablecaption{Derived Physical Properties for RRc Variables}
\tablehead{\colhead{ID}& \colhead{${\rm [Fe/H]_{ZW84}}$}& \colhead{$\langle M_{V}\rangle$} & \colhead{$M/\msun$}& \colhead{$\log(L/\lsun)$}&\colhead{$\log T_{\rm eff}$}& \colhead{Y}}
\startdata
V10& -1.810&  0.677& 	0.593& 	1.738&  3.860& 	0.266\\
V12*& -1.980& 0.782&    0.893&  1.774&  3.860&  0.245\\
V25& -1.710&  0.682& 	0.604& 	1.720&  3.863& 	0.270\\
V28& -1.722&  0.709& 	0.631& 	1.720&  3.863& 	0.269\\
\hline
\rm{Mean}& -1.747$\pm$0.032 & 0.689$\pm$0.010 & 0.609$\pm$0.011 & 1.726$\pm$0.006 & 3.862$\pm$0.001 & 0.268$\pm$0.001\\
\enddata
\tablecomments{Due to the unusual values obtained for V$12$ it is not included when determining the mean values for the physical properties of the RRc stars.}
\label{retcphysical}
\end{deluxetable*}

%The RRc star V$12$ is much bluer than the other RRc stars in Reticulum, being located near the blue edge of the instability strip, and has a much smaller amplitude, which is consistent with its blue color.  The physical properties obtained from the Fourier decomposition for V$12$ also show a marked difference from those of the other RRc's; we obtained a metalicity of ${\rm [Fe/H]_{ZW84}}=-1.98$, more metal poor than the other RRcs, and a mass of $M/\msun=0.89$ which is very large for an RRc.  Due to unusual values obtained for V$12$, it has not been included when determining the mean values of the physical properties for the RRc stars.  The very blue color of V$12$ and its small amplitude suggest that it could possibly be blended with a faint companion, however the $V$-band amplitude is also smaller than for the other RRc stars but V$12$ shows no sign of being brighter in $V$ than the other RR Lyrae.

\section{Distance Modulus}

The absolute magnitude-metallicity relationship from \citet{catelancortes08} is used to provide the absolute magnitude of the RR Lyrae stars in Reticulum.  The disagreement between the metallicities for RRab and RRc stars raises an issue as to which metallicity to use for calculating the absolute magnitude.  Since the metallicity obtained from the RRab stars is consistent with what has been reported in the literature and is drawn from a larger number of stars, that value is used, ${\rm [Fe/H]_{ZW84}}=-1.61\pm0.02$.  This value gives an absolute magnitude of $M_{V}=0.61\pm0.20$.  The average magnitude of the RRab stars, $\langle V\rangle=19.064\pm0.008$, and the reddening value of $E(B-V) = 0.016$ from \citet{schlegel98} are used, along with a standard extinction law of $A_{V}/E(B-V)=3.1$, to obtain a reddening-corrected distance modulus of $(m-M)_{0}=18.40\pm0.20$, which agrees with the value of $18.39\pm0.12$ found by \citet{ripepi04}.  This is shorter than the distance modulus of $(m-M)_{LMC}=18.44\pm0.11$ that \citet{catelancortes08} derived for the LMC, though the two distance moduli agree within the errors.  This is not necessarily a surprise as Reticulum is widely separated from the disk of the LMC, having a location that is about $11$ degrees from the center of the LMC \citep{walker92a}.

We can also use the period-metallicity-luminosity relationship for the $I$-band from \citet{catelan04} to determine the distance modulus.  We used the Fourier derived individual metallicities for the stars that we were able to successfully fit; we used the average metallicity of ${\rm [Fe/H]_{ZW84}}=-1.61$ for the remaining stars.  Using the $E(B-V) = 0.016$ from \citet{schlegel98} and a standard extinction law of $A_{I}/A_{V}=0.482$ gives an $I$-band extinction of $A_{I}=0.024$.  This gives a reddening-corrected distance modulus of $(m-M)_{0}=18.47\pm0.06$ which is longer than value obtained from the $V$-band magnitudes; the smaller error bar for the $I$-band based distance modulus is due to there being no systematic zero point uncertainty in the $I$-band.  

Despite the changes in color during the pulsation cycle of RR Lyrae stars, RRab stars show a very small range of intrinsic $B-V$ and $V-I$ colors during their minimum light phase \citep{mateo95}.  We compare the $B-V$ and $V-I$ colors of our RRab stars to the expected colors in order to calculate the reddening to the cluster.   We calculate the expected $B-V$ colors of the RRab stars using the method devised by \citet{sturch66} which gives the expected color as a function of period and metallicity.  We use the calibration of Sturch's method from \citet{walker92b} which gives the color excess as
\begin{equation}
E(B-V) = (B-V)_{min}-0.336-0.24P-0.056{\rm [Fe/H]_{ZW84}}
\end{equation}
where $P$ is the period of the star in days.  Table \ref{vimintable} lists the $(B-V)_{min}$ colors and the reddening, $E(B-V)$ for each of the RRab stars in Reticulum.  The Fourier derived individual metallicities were used for the stars for which they were successfully determining, we used the average cluster metallicity of ${\rm [Fe/H]_{ZW84}}=-1.61$ for the remaining RRab stars.  The average reddening is $E(B-V)=0.05\pm0.01$ which is larger than the $E(B-V)=0.016$ from \citet{schlegel98} but in agreement with the value of $E(B-V)=0.05\pm0.02$ found by \citet{walker92a} using the same method.  Unlike the expected $B-V$ color, the expected $V-I$ color at minimum light does not appear to vary significantly with the period or metallicity of the RRab and we use the value of $(V-I)_{0,min}=0.58\pm0.02$ from \citet{guldenschuh05}; Table \ref{vimintable} lists the $(V-I)_{min}$ colors and the reddening, $E(V-I)$ for each of the RRab stars in Reticulum.  The average reddening from the RRab stars is $E(V-I)=-0.01\pm0.02$; this is less than the reddening value of $E(V-I)=0.026$ that is expected based on the reddening from the Schlegel dust maps.

Table \ref{vimintable} shows that V$06$ has an $E(B-V)$ that is significantly larger than the values for any of the other RRab stars and an $E(V-I)$ value that is one the smallest.  V$06$ potentially shows the Blazhko effect and the light curve modulations that result from this effect could potentially make impact its color at minimum light.  If we exclude V$06$ from our calculations, we obtain an $E(B-V)=0.04\pm0.01$ and an $E(V-I)=-0.01\pm0.01$; moving both values closer to what is expected from the Schlegel values.

\begin{deluxetable}{lcccc}
\tablewidth{0pc}
\tabletypesize{\scriptsize}
\tablecaption{Reddening from RRab Stars}
\tablehead{\colhead{ID}& \colhead{$(B-v)_{min}$}& \colhead{$E(B-V)$}& \colhead{$(V-I)_{min}$}& \colhead{$E(V-I)$}}
\startdata
V01& 0.35& -0.01& 0.52& -0.06\\
V02& 0.46& 0.07&  0.61& 0.03\\
V05& 0.45& 0.07&  0.59& 0.01\\
V06& 0.51& 0.12&  0.51& -0.07\\
V07& 0.42& 0.05&  0.52& -0.06\\
V08& 0.44& 0.03&  0.55& -0.03\\
V09& 0.39& 0.01&  0.56& -0.02\\
V13& 0.44& 0.05&  0.63& 0.05\\
V14& 0.43& 0.05&  0.58& 0.00\\
V16& 0.44& 0.07&  0.58& 0.00\\
V17& 0.44& 0.08&  0.54& -0.04\\
V18& 0.42& 0.04&  0.59& 0.01\\
V19& 0.39& 0.03&  0.58& 0.00\\
V20& 0.40& 0.02&  0.57& -0.01\\
V21& 0.44& 0.04&  0.57& -0.01\\
V22& 0.42& 0.05&  0.59& 0.01\\
V23& 0.35& 0.00&  0.60& 0.02\\
V26& 0.47& 0.06&  0.60& 0.02\\
V27& 0.40& 0.03&  0.51& -0.07\\
V29& 0.44& 0.08&  0.60& 0.02\\
V30& 0.43& 0.05&  0.61& 0.03\\
V31& 0.38& 0.01&  0.55& -0.03\\
\enddata
\label{vimintable}
\end{deluxetable}

\section{The CMD}

Our color-magnitude diagram (CMD) is compared to theoretical isochrones from the Princeton-Goddard-PUC (PGPUC) stellar evolutionary code \citep{valcarce12}.  The RR Lyrae distance modulus of $(m-M)_V = 18.45$ mag is adopted; assuming a reddening of E(B-V)=$0.016$ \citet{schlegel98}, the true distance modulus is $(m-M)_{V,0} = 18.40$ mag, as discussed in \S5.  Although there have been some suggestions that the reddening towards Reticulum is larger than that adopted here (e.g., Mackey \& Gilmore 2004), as shown below, we find little evidence to support a larger reddening value than reported by \citet{schlegel98}; the reddening value obtained from the $B-V$ colors of the full sample of the RRab stars at minimum light does support the larger reddening value of Mackey \& Gilmore but the $V-I$ colors do not support such a large reddening.  Isochrones with a "normal" $\rm[\alpha/Fe]$ ratio (e.g., Mateluna et al. 2012), $Y=0.245$ and Z=$0.0006$ (corresponding to ${\rm[Fe/H]_{UVES}}\sim-1.61$ dex) are over-plotted.  

\begin{figure}
\epsscale{0.9}
\plotone{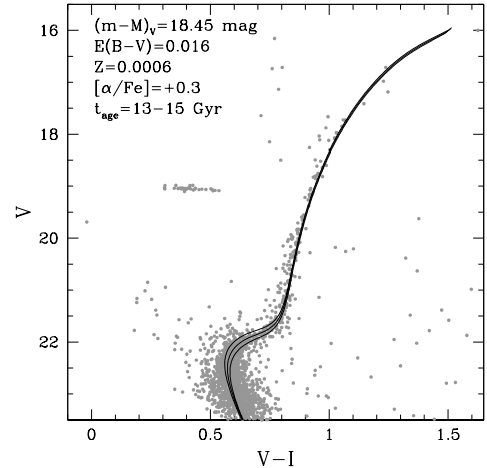}
\caption{The $V$,$\rm(V-I)$ CMD for Reticulum with theoretical isochrones from the Princeton-Goddard-PUC stellar evolutionary code \citep{valcarce12} also plotted.  The isochrones fit the observed RGB well and suggest an age of $\sim14$ Gyr for Reticulum.}
\label{ageplot}
\end{figure}

Figure \ref{ageplot} shows that the best fit isochrones have ages of $\sim14\pm2$ Gyr, consistent with the ages of other LMC GCs (Olsen 1999, Mackey \& Gilmore 2004, Bekki et al. 2008).  The observed RGB fits the CMD well.  In contrast, a larger reddening value would shift the isochrones to the red.  A smaller $\rm [\alpha/Fe]$ or a more metal-poor [Fe/H] would shift the isochrones to the blue, as would a smaller reddening value.  We therefore see no need to adopt a larger value of reddening than that that found by Schlegel et al. (1998).  A small reddening value is also in agreement with the E(B-V)=$0.03$ derived by \citet{walker92a}.  We believe Figure \ref{ageplot} provides evidence that our derived RR Lyrae distance modulus fits the CMD remarkably well and supports an old age of Reticulum.  

\begin{figure}
\epsscale{0.9}
\plotone{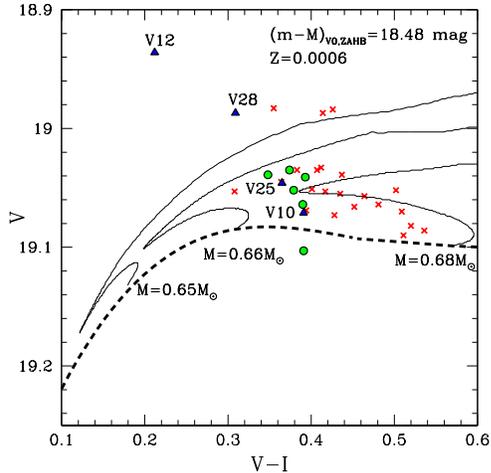}
\caption{The V,(V-I) CMD for Reticulum centered on the RR Lyrae instability strip.  A Zero Age Horizontal Branch from the BaSTI HB tracks (Pietrinferni et al. 2004, 2006) with Z=$0.0006$ is over-plotted, as well as  the BaSTI evolutionary tracks for $0.65\msun$, to $0.68\msun$ HB stars.  RRab stars are indicated by red X's, RRc stars by blue triangles, and RRd stars by green circles.}
\label{iso}
\end{figure}

Figure \ref{iso} shows the $V$,$\rm (V-I)$ CMD for Reticulum centered on the RR Lyrae instability strip.  A Zero Age Horizontal Branch (ZAHB) from the BaSTI HB tracks (Pietrinferni et al. 2004, 2006) with Z=$0.0006$ is over-plotted, as well as  the BaSTI evolutionary tracks for $0.65\msun$, to $0.68\msun$ HB stars.  As shown by Gallart et al. 2005, the deviation of the mean RR Lyrae magnitudes from the ZAHB is $\delta(V_{ZAHB}-<V>)_{RR} \sim 0.1$ mag at the metallicity of Reticulum, and hence we adopt $(m-M)_{V0,ZAHB} = 18.50$ mag.  The BaSTI tracks indicate that most of the Reticulum RR Lyrae stars have a mass range of $0.65 - 0.68\msun$.  These RR Lyrae masses are a little larger than those found from the Fourier decomposition of the RRc stars (see Table \ref{retcphysical}), although the mass of V28 derived from the BaSTI tracks and from the Fourier decomposition technique agrees remarkably well.  We note that changing $(m-M)_{V0,ZAHB}$ does not affect the derived RR Lyrae masses. In contrast, a change in Z affects the theoretical RR Lyrae masses in a sense that a more metal-rich Z shifts the RR Lyrae masses to smaller values.

\section{Oosterhoff Classification}

\begin{figure*}
\includegraphics[width=0.49\textwidth]{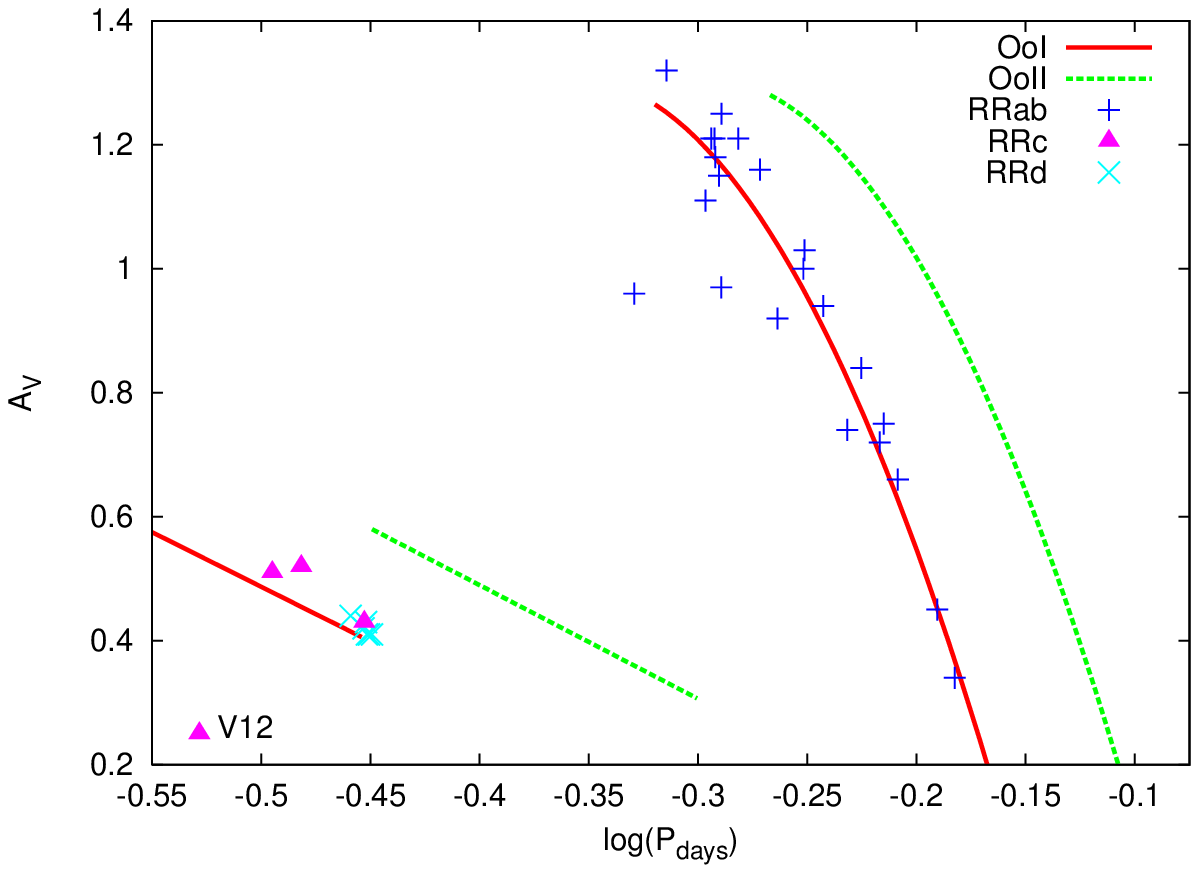}
\includegraphics[width=0.49\textwidth]{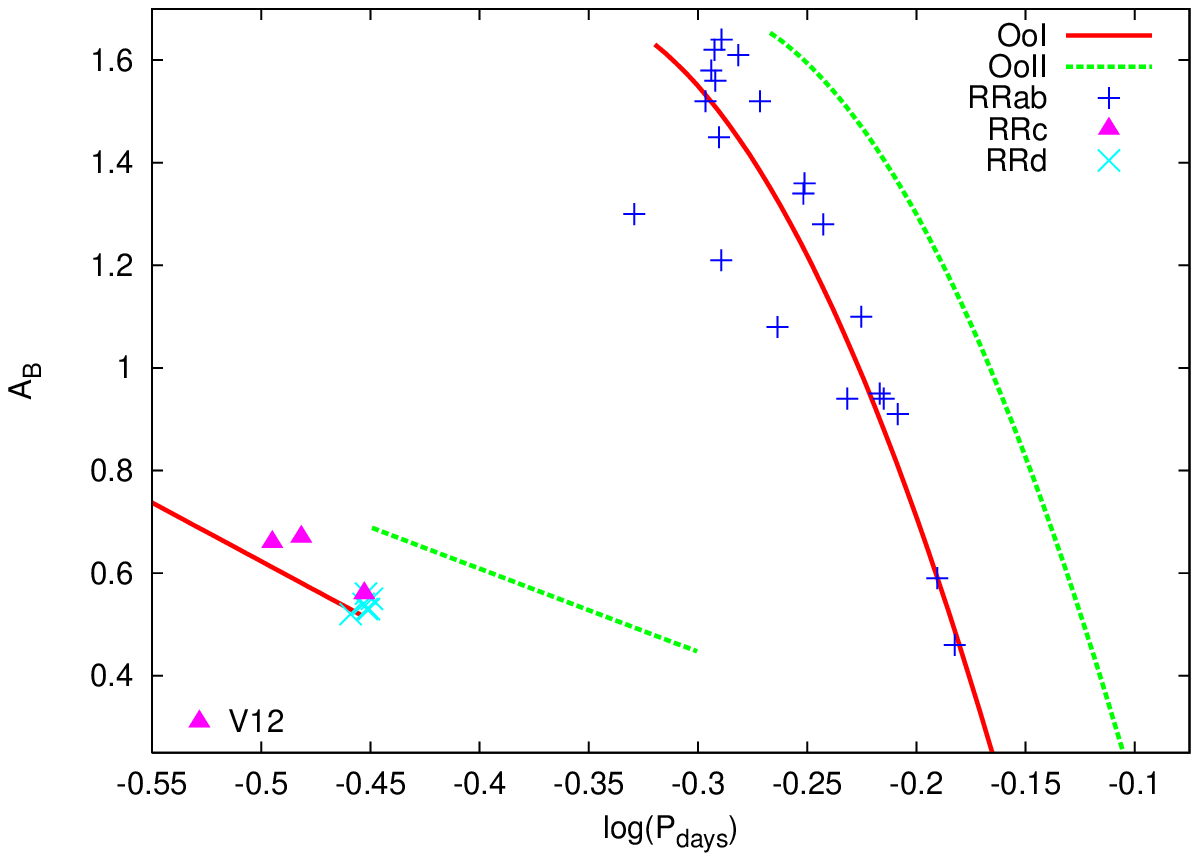}
\includegraphics[width=0.49\textwidth]{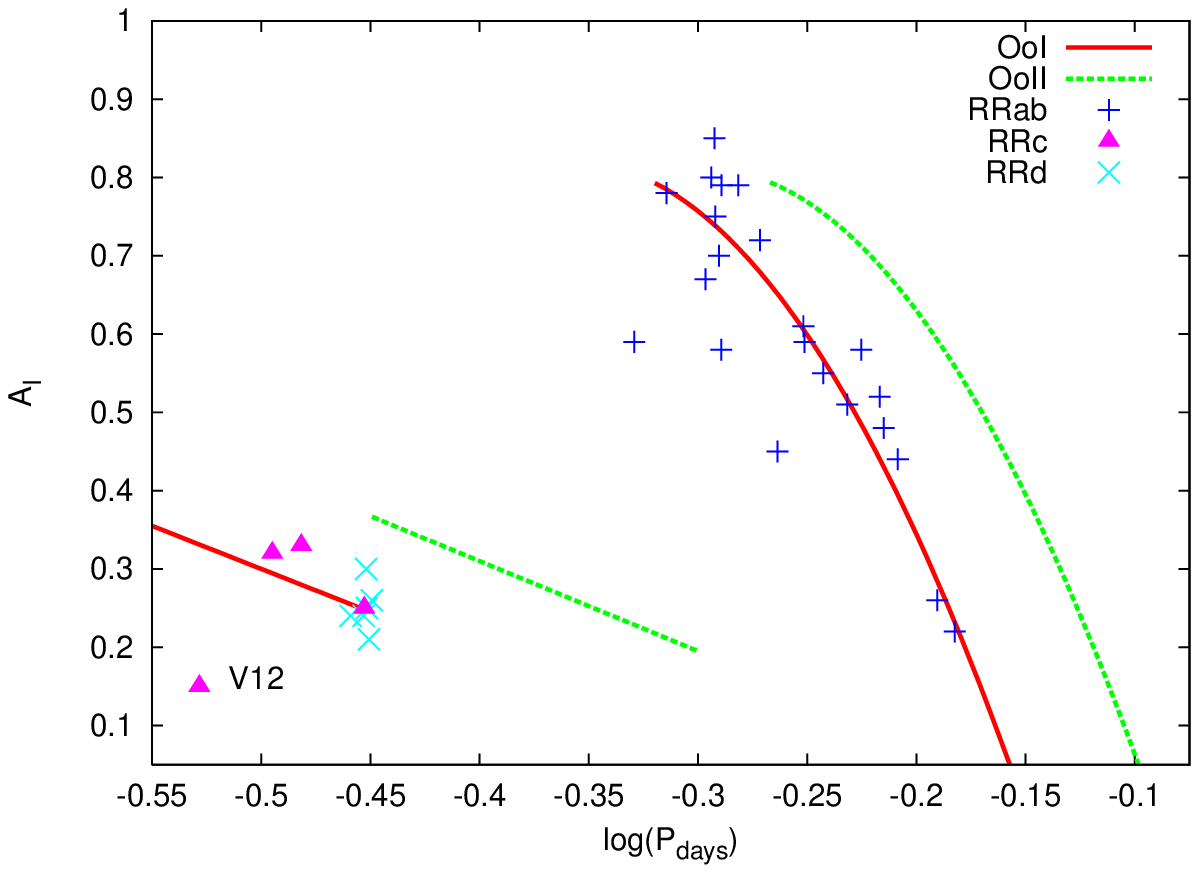}
\caption{Bailey diagrams, log period vs $V$-band(top left), $B$-band (top right), $I$-band (bottom) amplitude for the RR Lyrae stars in Reticulum.  Red and green lines indicate the typical position for RR Lyrae stars in Oosterhoff I and Oosterhoff II clusters, respectively \citep{cacciari2005,zorotovic10,kunder12}.}
\label{retvperamp}
\end{figure*}

The average periods for the RR Lyrae stars in Reticulum are $\langle P_{ab}\rangle=0.552$ days and $\langle P_{c}\rangle=0.325$ days.  The $22$ RRab stars, $4$ RRc's, and $6$ RRd's give the cluster a number fraction of $N_{c+d}/N_{c+d+ab} = 0.31$.  The average periods for the RRab and RRc stars strongly indicate an Oosterhoff I classification and, while the number fraction is high, it is still consistent with the cluster being an Oo-I object.  The minimum period for an RRab star in Reticulum is $P_{ab,min}=0.46862$ days, which is also consistent with an Oo-I classification for Reticulum \citep{catelan12}.

Figure \ref{retvperamp} shows the $V$, $B$, and $I$ band period-amplitude diagrams for Reticulum.  Both diagrams show that the RRab stars cluster along the line that indicates the typical location for RRab stars in Oo-I clusters.  There is more scatter in the positions of the RRc stars but most of them still are located near the Oo-I locus, confirming the classification of Reticulum as an Oo-I object.

\section{Conclusion}

We have conducted a photometric study of the Reticulum globular cluster in order to identify and classify the variable stars in that cluster; our data set consists of $228$ $V$, $222$ $B$, and $80$ $I$ images, making it the largest such data set on Reticulum.  We found a total of $32$ RR Lyrae stars ($22$ RRab, $4$ RRc, and $6$ RRd) in the cluster.  While all $32$ stars had been previously discovered, we were able to discover secondary pulsation periods in $2$ stars that had previously been classified as RRc stars.

We calculated Fourier parameters for a sub-sample of the RRab and RRc stars and used these to determine the physical properties of the RR Lyrae stars in Reticulum for the first time.  A future paper in this series will compare these physical properties to those obtained for other clusters in order to look at the differences between clusters of different Oosterhoff type.

We calculated a reddening-corrected distance modulus of $(m-M)_{0}=18.40\pm0.20$ which agrees with the literature values for Reticulum.

The $V,(V-I)$ CMD of the cluster was used to calculate an age of $\sim14\pm2$ Gyr for Reticulum, consistent with the age of the other old globular clusters in the LMC.  The CMD, along with the $V-I$ colors of the RRab stars at minimum light, do not support the suggestions that the reddening toward Reticulum is larger than the value of $E(B-V)=0.016$ from \citet{schlegel98}; however, the $B-V$ colors of the RR Lyrae at minimum light support the larger reddening value of $E(B-V)=0.04\pm0.01$ from \citet{mackeygilmore04}.

The average periods for the RRab and RRc stars indicate that Reticulum is an Oosterhoff I cluster.  This is confirmed by the location of the RRab and RRc stars on the Bailey diagram and the location of the RRd stars on the Petersen diagram.

\section{Acknowledgments}
Support for H.A.S. and C.A.K. is provided by NSF grants AST 0607249 and AST 0707756.  M.C. and J.B. are supported by the Chilean Ministry for the Economy, Development, and Tourism's Programa Iniciativa Cient\'{i}fica Milenio through grant P07-021-F, awarded to The Milky Way Millennium Nucleus, and by the BASAL Center for Astrophysics and Associated Technologies (PFB-06).  M.C. is also supported by Proyecto Fondecyt Regular \#1110326 and by Proyecto Anillo ACT-86.  J.B. is also supported by Proyecto Fondecyt Regular \#1120601.  We would like to thank an anonymous referee for helpful comments which improved this paper.

\end{document}